\documentclass[12pt]{article}
\usepackage{booktabs} 
\usepackage{paralist}
\usepackage{multirow}
\usepackage{amssymb}
\usepackage{todonotes}
\usepackage{xcolor}
\usepackage{url}
\usepackage{graphicx}
\usepackage{float}

\usepackage{fullpage}
\newif\ifshowcomments
\showcommentstrue
\ifshowcomments

\newcommand{\mynote}[2]{\fbox{\bfseries\sffamily\scriptsize{#1}}
 {\small$\blacktriangleright$\textsf{\emph{#2}}$\blacktriangleleft$}}
\else
\newcommand{\mynote}[2]{}
\fi
\newcommand{\ie}{i.\,e., }
\newcommand{\eg}{e.\,g., }

\begin{document}
\title{Building an Emulation Environment for Cyber Security Analyses of Complex Networked Systems}


\author{Florin Dragos Tanasache$^{1}$, Mara Sorella$^{1}$, Silvia Bonomi$^{1, 2}$,\\ Raniero Rapone, Davide Meacci\\
$^{1}$DIAG - Sapienza University of Rome, Via Ariosto 25, 00185, Rome, Italy\\
$^{2}$CINI Cyber Security National Laboratory, Via Ariosto 25, 00185, Rome, Italy\\
\texttt{tanasache.1524243}@studenti.uniroma1.it,\\ \texttt{\{sorella,bonomi\}}@diag.uniroma1.it, \\
\texttt{raniero.rapone}@outlook.com, \texttt{davide.meacci}@gmail.com
}

\date{}
\maketitle

\begin{abstract}
Computer networks are undergoing a phenomenal growth, driven by the rapidly increasing number of nodes constituting the networks. At the same time, the number of security threats on Internet and intranet networks is constantly growing, and the testing and experimentation of cyber defense solutions requires the availability of separate, test environments that best emulate the complexity of a real system.
Such environments support the deployment and monitoring of complex mission-driven network scenarios, thus enabling the study of cyber defense strategies under real and controllable traffic and attack scenarios.

\noindent In this paper, we propose a methodology that makes use of a combination of techniques of network and security assessment, and the use of cloud technologies to build an emulation environment with adjustable degree of affinity with respect to actual reference networks or planned systems. As a byproduct, starting from a specific study case, we collected a dataset consisting of complete network traces comprising benign and malicious traffic, which is feature-rich and publicly available.
\end{abstract}

\newpage

\newcommand{\silvia}[1]{\textcolor{red}{\mynote{SB}{#1}}}
\newcommand{\mara}[1]{\textcolor{blue}{\mynote{MS}{#1}}}
\newcommand{\sncs}[1]{\textcolor{green}{\mynote{FT}{ #1}}}

\newcommand\temporary[1]{{\color{gray}#1}}
\newcommand\maraedit[1]{{\color{blue}#1}}
\newcommand\florinedit[1]{{\color{green}#1}}
\newcommand\silviaedit[1]{{\color{red}#1}}


%
%
%

\section{Introduction}

\noindent {\bf Context and Motivation.} Starting from the past decade, cyber attacks have become increasingly sophisticated, stealthy, targeted and multi-faceted, featuring zero-day exploits and highly creative interdisciplinary attack methods.
Furthermore, it is most likely an upward trend, due to two main factors: (i) increasing complexity of ICT networks and (ii) increasing capabilities of cyber attackers.
As a consequence, cyber security is becoming of primary importance for every organization, including small and medium enterprises and large companies, as well as critical infrastructures on which our society is becoming increasingly dependent.

One of the countermeasures that can be taken to face attackers is to try to play the role of the attacker and stress the environment that is aimed to protect. However, in order to enable this possibility, a separate, dedicated emulation/simulation environment must be set in place. 
Such environment should be able to: (i) represents realistic cyber environments that fit the testing objectives, (ii) offer tools for producing both benign and malicious system events and (iii) support the definition and creation of new scenarios and cyber threats in a cost and time-effective manner~\cite{eskridge2015vine,pham2016cyris}.

Combining network and security assessment techniques with the use of cloud technologies allows for the deployment of emulation environments.
The objective of an emulation environment is enabling the deployment of fully virtualized instances of computer networks, with adjustable degree of affinity with respect to an actual reference networks or planned systems~\cite{raniero18}, allowing for arbitrary speculative analyses, notably for what concerns cyber security.\\

\noindent{\bf Related Work.} 
To the best of our knowledge, currently, very few complete solutions exist to support the design, development and deployment of an emulation environment to support testing and training.
In~\cite{eskridge2015vine} the authors proposed a Virtual Infrastructure for Network Emulation (VINE), pointing out for a set of tools that helps the construction of virtual topologies allowing for different type of experiments for cyber security researchers. 
Furthermore, VINE supports traffic behavior generation using monitored agents or the traditional packet capture for a low-level analysis of the state of the hosts. Operators can exploit a web-based interface, to manually create, modify, and monitor experiment testbeds. 
The main drawback of this paper is that the platform is overviewed at a rather general viewpoint, without allowing for a deep comparison.
In the specific context of cybersecurity, in ~\cite{pham2016cyris}, the Cyber Range Instantiation System (CyRIS) system is presented, as a solution, based on OpenStack, for reliably and repeatably deploying cyber ranges, \ie, technology development environments for cybersecurity training courses. 
It supports basic automation for systems and network deployment, and a certain degree of automation for sophisticated security activities (such as modifying firewall rules or planting malware). Another, close-sourced example is~\cite{yasuda2016alfons}. 
While these works provide valid tools for deploying testbeds for cyber experimentation, none of them tackles the task of documenting a methodological framework for creating emulation environments (along with the multiple existing technical tradeoffs).
Our paper, instead, presents a complete end-to-end methodology to build a virtual environment and deploying a cyber security testbed, mapped from an existing network. Furthermore, our work directly shows the benefit of the approach by providing a realistic dataset which can benefit multiple sides of the cyber-security scientific community.\\

\noindent {\bf Contributions.}
This paper provides three main contributions: (i) it proposes a methodology for designing and deploying a cloud-based emulation environment, (ii) it discusses the application of the proposed methodology to a case study and (iii) it shows the flexibility and power of the proposed approach to build an environment that is able to generate a realistic dataset, comprising of benign and malicious traffic. Furthermore, the dataset has been processed to obtain feature-rich labeled attack flows, useful for cyber security analyses such as training IDS and IPS classifiers and other machine-learning tasks, as well as for deep packet inspection investigation and related activities.
In particular, we exploit our practical study case to highlight, for each step of the methodology, how the relevant technologies and tools can be combined to achieve specific design requirements in an effective way.



\section{Methodology}
\label{sec:methodology}

In this section, we present a methodology for the design and deployment of a cloud-based emulation environment to support cyber-security testing, analysis and training activities.\\

\noindent The main goal of our methodology is to drive designers and system analysts in the creation of a virtualized environment that is able to reproduce a target networked physical environment, which will be used as a testbed.
The virtual environment will be deployed in a private cloud (leveraging IaaS frameworks) built on dedicated hardware.
Our approach is articulated in three main activities: (i) \emph{Virtual Environment Infrastructure setup} (\ie how to set up a private cloud based on the physical infrastructure), (ii) \emph{Virtual Network Design} and (iii) \emph{Data Collection} from a deployed testbed.
We will use the term \emph{virtual} and \emph{emulation} environments interchangeably to refer to the enabling infrastructure and emulation/virtual \emph{testbed} to mean the reproduced network within it.\\

\noindent {\bf Virtual Environment Infrastructure Setup.} This activity takes care of the design and deployment of the virtual environment on dedicated hardware. 
It is articulated along the following tasks: (a) select a cloud management platform to manage the storage and compute resources provided by the physical infrastructure, 
(b) define storage management policies for the virtual machines templates repository, in a way to foster data locality at instantiation time and fault tolerance and finally 
(c) design and deployment of the virtual communication infrastructure. 
Section \ref{sec:meth_env} discusses this activity, suggesting technological options to support each task.\\

\noindent {\bf Virtual Network Design} This activity has the objective, starting from a reference network specification, to create a testbed configuration, \ie all the metadata needed to deploy its virtual counterpart. 
This can be done main in two ways: (i) by manually defining the characteristics of the required network or (ii) by gathering information from an existing network in order to clone it or use it as starting point. In the following, we will consider the second and most challenging case. 
To this aim, we will discuss the following necessary tasks: (i) network topology identification,(ii) OS detection, and (iii) active service discovery in Section~\ref{sec:meth_ref_mapping}.\\

\noindent {\bf Data Collection} This activity is responsible of ensuring proper handling of the experimental data (in the form of traffic captures) generated by the emulated testbed, 
and is reported in Section~\ref{sec:meth_data_coll}.



\subsection{Virtual Environment Infrastructure}
\label{sec:meth_env}
\begin{figure*}
	\centering 
	\includegraphics[width=1\linewidth,trim={1cm 0cm 0.5cm 2cm},clip]{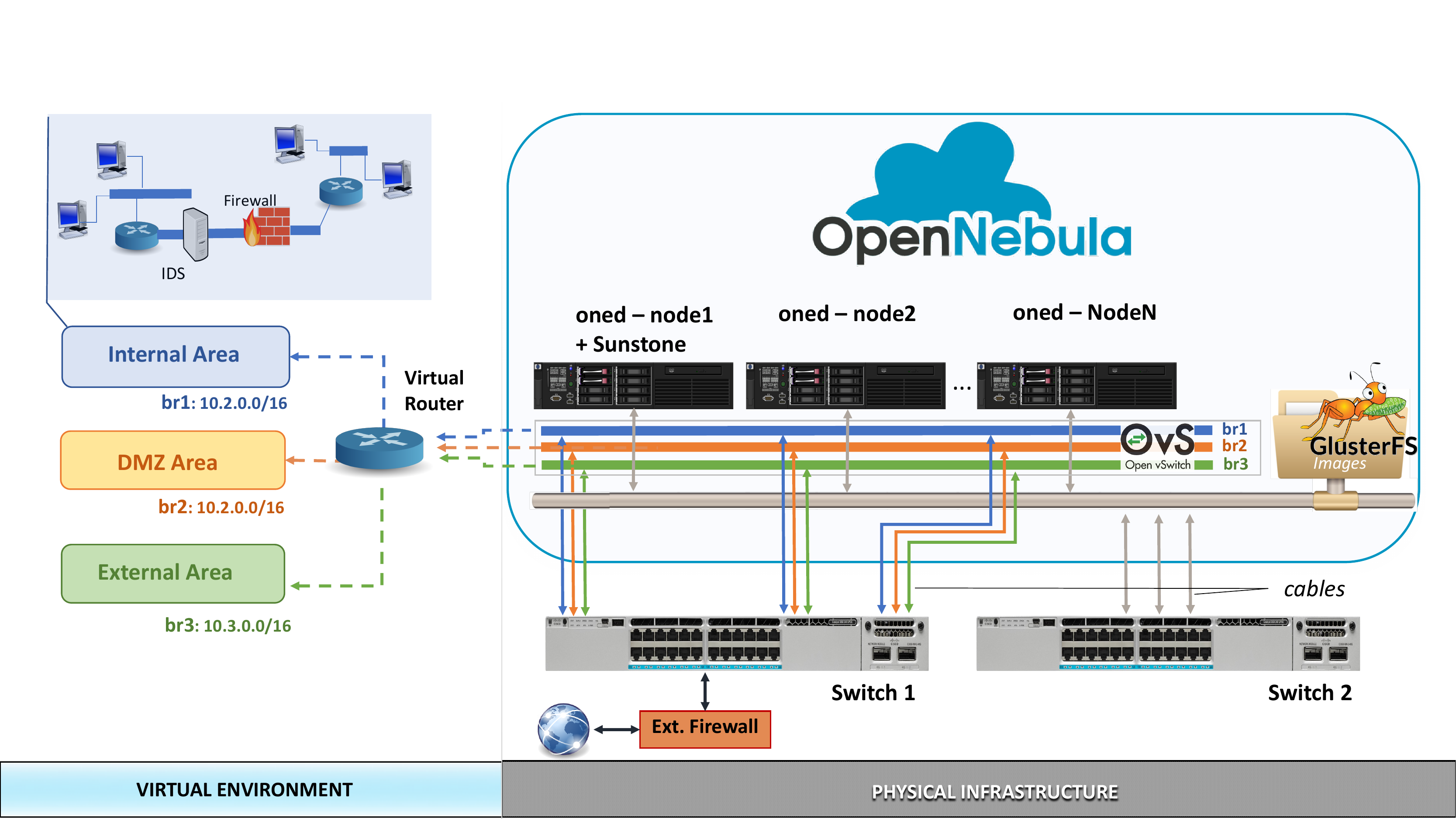}
	\caption{Overview of the Emulation Environment infrastructure. An example configuration of the communication layer is reported, making use of three OVS bridges, namely \texttt{br1} (internal lans), \texttt{br2} (DMZ service lans) and \texttt{br3} (external lans).}
	\label{fig:meth_env_environment}	
\end{figure*}

In the following, we provide an overview of the main steps required to build up an emulation environment stack that fits our purposes of network emulation and cyber security experimentation.


\subsubsection{Private cloud Setup}

Many IaaS platforms exist, and each one has its specific characteristics, furthermore, some platforms are constantly developing. 
The two major open-source ones by market penetration are OpenStack and OpenNebula, currently widely used by many people around the world~\cite{von2012comparison,vogel2016private}.
In order to set up our private cloud environment, we compared them taking into account the following features, namely \emph{Internal Organization, Software Deployment, Storage, Networking, Hypervisors and Governance}.
Due to lack of space, the detailed comparison with respect to the selected features is reported in Appendix~\ref{app:iaas_comp}.
After the analysis and experimentation with both, we found OpenNebula's solution both enterprise-ready and easy to use and deploy, thus being a good fit for any organization.  We chose OpenNebula as our IaaS for our case study (Section~\ref{sec:case_study}).

\subsubsection{Storage and Distributed File System}
\label{sec:meth_env_dfs}
In a typical real environment, machines run various operating systems. 
On all cloud platforms, VM types have a corresponding notion of \emph{template}\footnote{Templates on OpenNebula documentation: \url{http://docs.opennebula.org/5.4/operation/vm_management/vm_templates.html}}: a collection of data (\ie a set of disk images) and metadata (\ie no. of associated cores, RAM, network interfaces, booting order, context information, etc).

Thus, typically when setting up an emulated environment, a proper template repository must be maintained, \ie a physical area where templates are stored, so that they can be easily browsed and recalled at VM instantiation time.
However, VMs can be instantiated on all physical nodes in the cluster, so a centralized repository would result in heavy load on the cluster (especially as such templates can be large, typically in the order of 20-100 Gbytes).
Therefore, a distribution or replication strategy could be implied.

GlusterFS (later, Gluster), is a scalable, distributed file system that aggregates disk storage resources from a pool of trusted storage servers into a single global namespace. It is used in production at thousands of enterprises spanning media, healthcare, government, education, web 2.0, and financial services~\cite{shirinbab2013performance}. A gluster \emph{volume} is a logical collection of export directories (called \emph{bricks}) located on the various servers of the pool. 
Various types of volumes are supported, and can be chosen depending on the distribution criteria that fits a specific use-case. To exploit data-locality at VM instantiation time, a good choice is setting up a \emph{Replicated GlusterFS Volume} 

Apart for the inherent data redundancy, this solution also offers reliability as in case of a brick failure, the data can still be recovered from its replicas. 
Gluster supports LVM Volumes as mount points, so that multiple disks can be used for each server and regarded as a single volume. 

Other options include Lustre and Ceph. 
OpenNebula storage is structured around the \emph{Datastore} concept, \ie any storage medium to store disk images.
It makes use of a Datastore called Images, that points to the GlusterFS mount point, plus, other two (non replicated) datastores: \emph{System},  containing the instantiated machines data and \emph{Files \& Kernels} to store plain text files such as scripts.

%
%

\subsubsection{Communication layer}
\label{sec:meth_env_comm}
In virtual environments, VMs are connected to virtual networks (VLANs) via virtual network interfaces. 

IntraLAN communication refers to interactions among VMs in the same VLAN, while interLAN refers to cross VLAN communication.
The VLAN deployment and communications are managed in first instance by the IaaS platform.
The Layer 3 communication between different VLANs is made possible by \emph{Virtual routers} (VR), another type of VM that works just like a physical Layer 3 router, thus allowing for the definition of specific routing tables, gateways, etc.. A VR has one network interface in each VLAN that it manages.
However VMs might be deployed on different physical servers, therefore VLANs can span different nodes, that are in turn connected via physical switches. 
\emph{Virtual switches} are a key component in virtual environments, as they provide Hypervisors with connectivity between virtual interfaces of the VMs and the physical interfaces of the various nodes. 
They are software modules running on all nodes that maintain a MAC addresses database, to keep track of the VM addresses in the various LANs, process incoming input frames from physical interfaces and forward them based on the database. 

For the intraLAN and interLAN communication the two most commonly used virtual switches are \emph{Linux Ethernet Bridge} and \emph{Open vSwitch}.\\\\
%
\textbf{Linux Bridge} is a layer 2 virtual device, integrated in Linux OS. It mainly consists of four major components: i) a set of managed physical interfaces, ii) a control plane managing the Spanning Tree Protocol (STP) (for network loops prevention), iii) a forwarding plane used to directs traffic to the appropriate interfaces based on mac address.\\\\
%
\textbf{Open vSwitch (OVS)} has been around since 2009 but hasn't been a contender to Linux Bridge before 2014. OVS provides the same functionality, plus other Layer3 functions. Licensed under the open Source Apache 2.0, it implements a virtual multilayer network switch, designed to enable effective network automation through programmatic extensions, while supporting standard management interfaces and protocols such as NetFlow, sFlow, SPAN, RSPAN, LACP and VLAN tagging.
OVS is used in multiple products and testing environments. 

\noindent Summing up the differences between the virtual switches, OVS is targeted at large multi-server virtualization environments, so it is focused on logical abstraction and management. The Linux bridge is fast and reliable, but lacks many control features.
IaaS like OpenStack and OpenNebula have started using OVS as default configuration in their networking framework. We also recommend the choice of OVS and assume this as a configuration choice in the following.

Multiple virtual switches (that in OVS are referred as \emph{bridges}) can be deployed to separate groups of VLANs logically, \eg internal lans versus others providing external services. 
Each VLAN is configured using the IaaS to sit on a single, specific bridge, as VMs receive their network address through the corresponding bridges (that, instead, have no address). 
\subsubsection{Physical Infrastructure Configuration}
\label{sec:meth_phys_infrastructure}
Typically, the required physical infrastructure for the deployment of a Virtual Environment infrastructure consists of a number of physical servers (or \emph{nodes}), each one with its own specifications, and a number of physical switches, used to connect them (Figure~\ref{fig:meth_env_environment}). 
The servers run the IaaS software: in the particular case of OpenNebula, a ``master'' node must be configured to host the OpenNebula daemon, \emph{oned}, and the Sunstone Frontend, that communicates with oned, via \emph{XML-RPC} APIs. Other nodes just need to have the Hypervisor installed (package \texttt{opennebula-kvm} on Debian based systems). 

If OpenNebula's Image Datastore is managed via GlusterFS, the servers must be connected through a switch (we will call this the ``backbone'' network): when new templates containing VM datastores are added to the master node, the GlusterFS daemon will start file transfers to keep replicas updated.
To speed the sync between the replicas the bandwidth must be maximized. This translates to using high-end network interfaces on both the servers and the switch (i.e., use at least 10Gb ports).
If this is not possible, a bandwidth improvement can be made via \emph{interface bonding}.\\\\
\textbf{Linux bonding} driver provides a method for aggregating multiple network interfaces into a single logical "bonded" interface. In fact, the bonding allows combines several network interfaces into a single link, providing either high-availability, load-balancing, maximum throughput, or a combination of these. There are several bonding modes. One of them is mode 0 (balance-rr). In this mode, packets are transmitted in sequential order on the bond interfaces.\\
Furthermore, to enable VM communication via OVS bridges, the servers must be connected to a switch using one physical port per bridge. It is a good practice to use a separate switch for the backbone network and for OVS bridges, in order to physically separate business traffic pertaining to the emulated environment, and communication between the replicas.\\
All the logical pieces comprising the Emulation environment are resumed in Figure~\ref{fig:meth_env_environment}.

\subsection{Virtual Network Design}
\label{sec:meth_ref_mapping}
As previously stated, one of the objectives of an emulation environment is obtaining a separate, virtual and arbitrarily modifiable copy of an existing network for conducting analysis and tests.
In order to do that, fundamental inputs are the Network and Vulnerability Inventories and communication rules.
Such inputs are commonly available in large enterprise networks as they are  outputs of a continuous monitoring process and can be used to trigger cyber attacks detection and response processes. It is important to note that the effectiveness of the monitoring, detection and response activities depends on the accuracy of the collected inventories.\\\\
Inventory collection tools, such as GFI LANGUARD Network Scanner, typically require an intrusive approach, where monitoring agents are preinstalled on each host system. Furthermore, such as for the case of LANGUARD, they usually require expensive licenses. 
For these reasons in the following we will direct our choices towards software tools that are free and allow to perform this phase in a non intrusive way.\\

Instead, when this information is not available, such inputs can, in principle, be collected manually by interviewing a network administrator that lists all the requirements and details of the desired virtual network or by collecting them starting from a real physical network.



\subsubsection{Network topology identification} This task allows to obtain a logic schema of the network infrastructure of the reference network in terms of ISO/OSI layers 2 and 3, i.e., of how many subnetworks it is composed, how they are separated using network devices (e.g., router, hubs, switches), as well as the presence of firewalls and their relative configuration rules.

Common approaches for network discovery rely on distributed traceroute monitors~\cite{donnet2005improved}: however, the accuracy of this approach strongly depends on the observability of the network.

Overall, this phase is the hardest to automate, as in order to be accurate it relies heavily on pre-existing knowledge about the reference network, such as the one coming from Network Administrators and Network schemes. 

In the following, we make the assumption that the information about the number and composition of network LANS, as well as the network devices, is known a priori or can be determined by querying the network administrator. 

\subsubsection{Host and OS Detection}
\label{sec:meth_ref_osdetection}
Once that the network topology is known, an important step in the reference network mapping concerns identifying the computer hosts deployed in the network, to be able to replicate their presence in the virtual testbed.  
To this end, the primary information required is discovering active hosts on the network and their OS.\\\\
\emph{OS detection} is the process of determining the OS of a remote computer~\cite{orebaugh2011Nmap}. When this is accomplished in a non intrusive way (\ie, without deploying agents), it is achieved by  observing the external behavior of the remote host, mostly in terms of network traffic, in a way to derive what is called an OS \emph{fingerprint}~\footnote{a 67-bit signature describing the host behavior.}, adopting either an \emph{active} or an \emph{passive} approach.

Active OS fingerprinting requires sending specially crafted packet probes to the system in question, aiming at eliciting a distinguishable responses from different OS'es, derived from the implementation dissimilarities of their communication protocols~\cite{medeiros2009data}. 
System responses from this active probing are then  used to generate the fingerprint and fed into a classification model.  Passive OS fingerprinting involves sniffing network traffic inside a network segment and matching known patterns to a table of pre-established OS identities. 
In general, active scanning allows for higher accuracy~\cite{richardson2010limits}, and passive is chosen only when a given degree of stealth or anonymity of capture is of concern
(e.g., when aiming at performing malicious activities).

\paragraph{Tools}
The most popular active network scanner is Nmap~\cite{lyon2009Nmap}
, an open source tool designed for network exploration and security auditing. Nmap uses raw IP packets in novel ways to determine what hosts are available on the network, what services (application name and version) those hosts are offering, what OS version they are running, what type of packet filters/firewalls are in use, and dozens of other characteristics. Its output is composed of a list of scanned targets together with supplemental information on each, depending on the options used. 
Another tool offering OS detection capability is OpenVAS, most known for its value as a vulnerability assessment tool (see Section~\ref{sec:meth_service_vuln}).
Let us note that each of these tools has its own characteristics and main target applications. As a consequence, the produced output will satisfy different level of quality depending on the considered attributes. To improve the overall quality (in terms of accuracy of the scan operation) it is mandatory to integrate several scanners combining their output. Due to the lack of space, we postpone to Appendix~\ref{app:hybrid_approach} our proposal for combining results coming from Nmap and OpenVAS.



\subsubsection{Active service discovery and vulnerability surface identification}
\label{sec:meth_service_vuln}
Determining what services are running on the network hosts is of primary importance to allow for an adequate mapping of the reference network.
In particular, for cyber security purposes, what is of interest is understanding the hosts' exposed ~\emph{vulnerability surface}, i.e. identifying which running services have an associated open port, and are thus, potentially directly exploitable.\\
A typical example is identifying DNS and HTTP servers are running and their respective versions. An accurate version number is key in determining which exploits a host is vulnerable to.\\
Vulnerability scanning tools usually produce a detailed report with the severity level of every vulnerability detected. 
\paragraph{Tools}
OpenVAS is the open-source spin-off of the popular security scanner Nessus, and provides a modular architecture composed of a scanning agent and a manager with a powerful interface. The security scanner, is accompanied with a daily updated feed of Network Vulnerability Tests (NVT). 
Being one of the most powerful free security scanners, supports thousands of vulnerabilities and offers false positive management of scanning results.
However, being a free tool it suffers from limitations with respect to its paid counterparts, such as Nessus, Nexpose or LANGUARD, such as limited DB, and cumbersome configuration.
Nmap is also capable of discovering open ports and associate services.

\subsubsection{Testbed Configuration}
\begin{figure}
	\centering 
	\includegraphics[width=\linewidth]{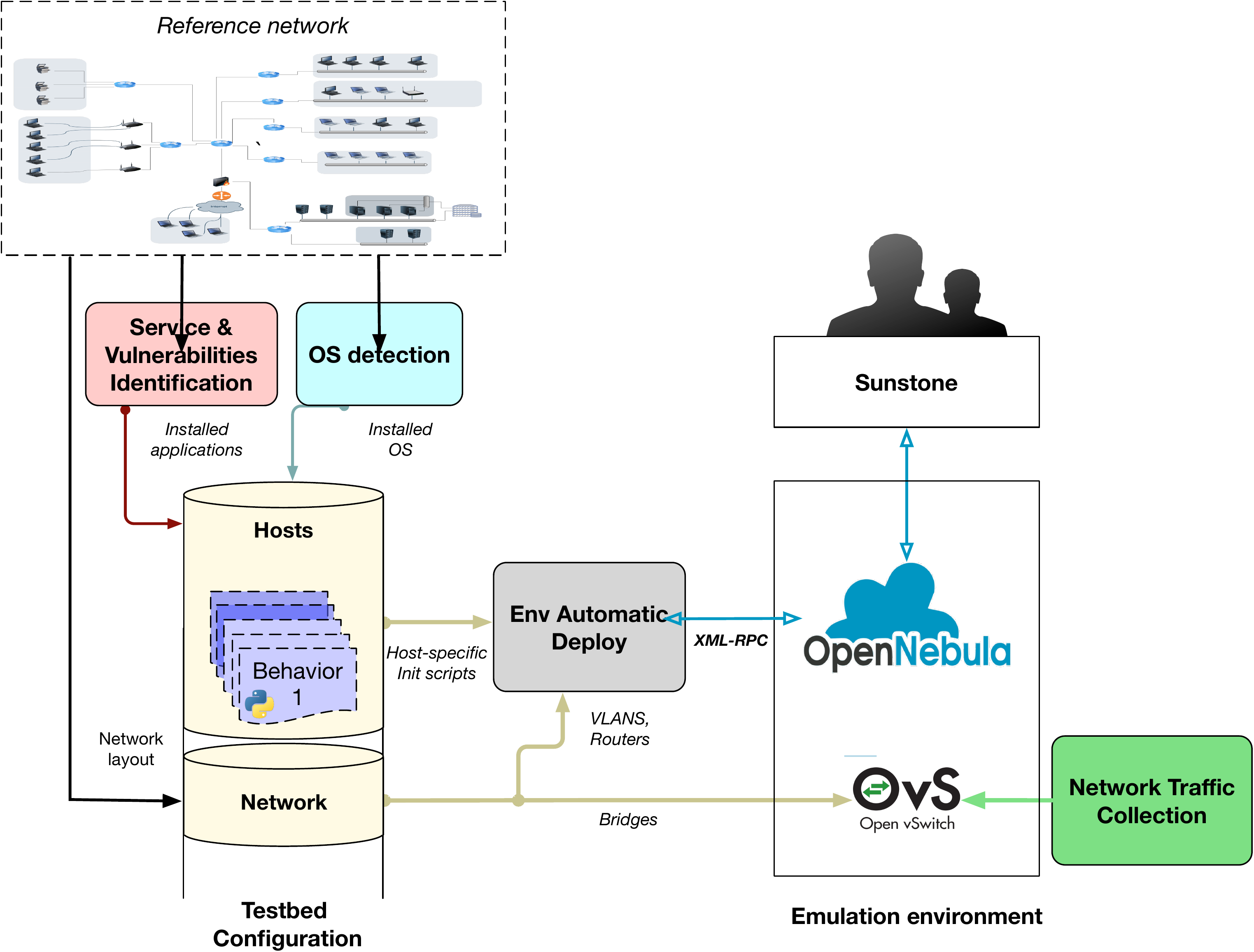}
	\caption{Overview of the emulation environment testbed deployment and experimentation phases.}
	\label{fig:system_overview}	
\end{figure}

In this section we summarize the metadata needed in order to deploy our testbed, derived from the phase of Virtual Network Design. It is composed of \emph{Hosts configuration} and \emph{Network configuration}, consolidated in the Testbed Configuration (Figure ~\ref{fig:system_overview}).\\\\
\textbf{Hosts Configuration} Each host identified from the reference network is associated to an OS template in the Images Datastore, a list of installed services CPE (with optional version) and a list of network interfaces. Optionally, a host-specific synthetic behavior can be embedded by deploying appropriate software code (see for example Section~\ref{sec:dataset_generators}).\\\\\textbf{Network Configuration} All network LANS will have a corresponding VLAN configuration, assigned to OVS Bridges also based on the data collection requirements (see Section~\ref{sec:meth_data_coll}). All layer 3 network devices will have a corresponding Virtual Router configuration, with their corresponding interfaces (see Section~\ref{sec:meth_env_comm}), physical Firewalls will correspond to Virtual Firewalls.

This information may be stored in a database or other means, if automation of the deployment is of concern. Most IaaS platforms offer their APIs to allow for automation of deployment to various extents. OpenNebula offers a centralized XML-RPC interface that can be used to entirely manage the emulated testbed in an automatic way, such as creating VLANs, instantiating VMs, etc. In Section~\ref{sec:case_study} can be found an example on how to fully automate the deployment of a test environment using OpenNebula API.

\subsection{Data Collection}
\label{sec:meth_data_coll}

To wrap up all information concerning the communication layer from Section~\ref{sec:meth_env_comm}, a testbed contains a number of virtual machines whose network interfaces belong to a number of VLANs sitting on a number of OVS bridges.
Layer3 routing between VLANs is handled by Virtual routers.In this section, we elaborate on different techniques and solutions for achieving network traffic data capture in an emulated testbed deployed in an emulation environment.\\\\
The problem of capturing and decoding traffic in high-speed network, be it physical or virtual has been deeply studied (see for example~\cite{braun2010comparing,fusco2010high}). In the context of a testbed for cyber security experimentation, the choice of a capturing solution depends on several criteria: reaction behavior, impact on network performance (delay, throughput), traffic rate and architectural considerations.\\\\
Typically, inline capture mode is preferred when the probe needs to react to intrusions (IPS) as it can alter or delete packets~\cite{kreibich2001network}. In particular, \emph{network taps} can be implied, \ie systems that are installed in network segments to monitor network devices such as 
routers, firewalls, servers and hosts by means of network, application, or security analyzers, achieving high-speed full duplex capture with no data loss.
Inline capture mode might alter network performances due to packet processing delays, yet the analysis capacity of Next-Generation Intrusion Systems (NGIPS) in inline transparent mode has now reached speeds higher than the physical link, allowing for extremely low latencies~\cite{meena2018integrated}.
However, inline taps are only suited to monitor a point-to-point link, thus  not being a viable choice when the full data capture is of concern.
Moreover, an inline probe failure can induce network failures.

On the other hand, passive mode is stealthier and does not alter the performances or the availability of the monitored network. 
A cheap option is capturing on the hardware switches, yet VM-to-VM traffic that does not pass over a physical wire (e.g., if they are on the same VLAN on the same server) would not be captured.

Depending on the capturing requirements, thanks to its high flexibility and capabilities OVS can come to rescue.

As mentioned in Section~\ref{sec:meth_env_comm}, OVS bridges can help distributing the VLANs in logic compartments; this choice can also be made with data collection purposes in mind:
indeed deploying the VLANs on different bridges allows the sniffing in specific parts of the environment. 
However, even if the physical interface associated to each OVS bridge could be monitored, the issues related on same-node traffic is still present. 

When internal traffic of the VLANs is of concern (as, for example, if full capture is desired), OVS supports port mirroring capabilities (with the SPAN and RSPAN protocol)~\cite{pettit2010virtual}. For each network to be monitored, SPAN allows to mirror the traffic from all network interfaces toward a specific output port (the SPAN port). To achieve full capture the best option is to have a SPAN port for each bridge, for each physical node. Span output ports can be for example a network interface of a ``Sniffer'' VM, where the traffic is then collected to be later analyzed.

However, this solution can potentially suffer from bandwidth limitation and data loss, as the full traffic is replicated and sent to the sniffer.

\section{A Case Study: DIAG Department}
\label{sec:case_study}
As a candidate for a case study, we chose the network of the Department of Computer, Control, and Management Engineering Antonio Ruberti (DIAG).
In particular, we followed the methodology described in Section~\ref{sec:methodology} and set up an emulation environment based on OpenNebula, using OVS for the connection layer, and GlusterFS for mantaining a replicated Images Datastore, and deployed an emulated testbed.

The network topology of the department consists of seven subnets, that we will call LAN1, LAN2, LAN3, LAN4, CED, DMZ\_INT and DMZ\_EXT.
The first three contain personal workstations (with a few exceptions of servers) of research staff belonging to various research areas, like Computer Science, Automation and Control and Management Engineering and the fourth pertains to administrative staff.
Furthermore, CED contains servers and compute nodes used by researchers to perform their work, while DMZ\_INT hosts internal services, such as SMB or SFTP and finally DMZ\_EXT hosts servers that expose services towards Internet (such as a WWW server and an SVN repository).

The following sections discuss each implementation step of the methodology in the context of this case study.
\subsection{OS Detection and Vulnerability Discovery}
\begin{figure}[H]
	\centering 
	\includegraphics[width=0.7\linewidth,trim={2.5cm 3.75cm 2.5cm 3cm},clip]{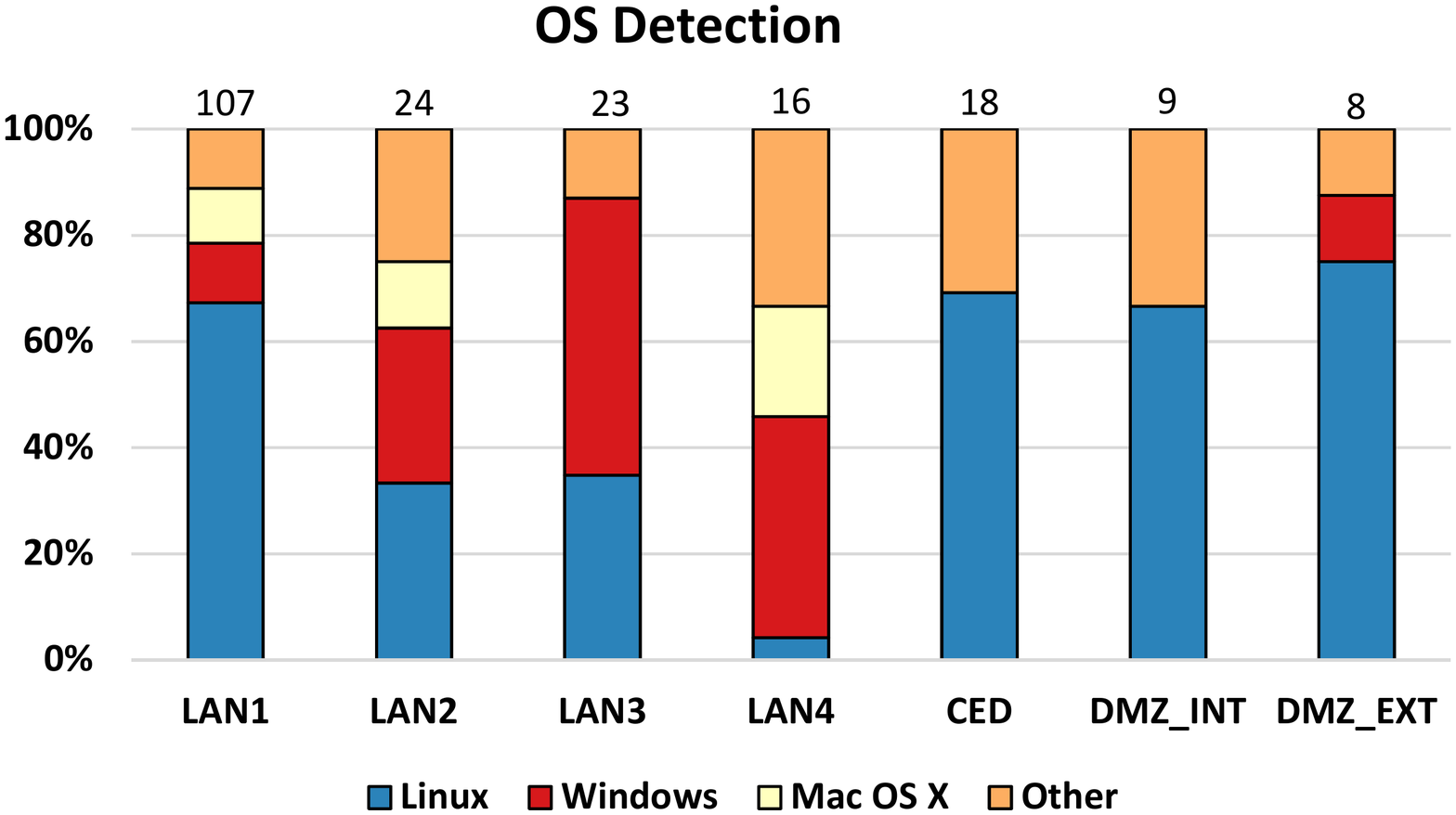}
	\caption{OS Detection results.}
	\label{fig:study_case_os_detection}
\end{figure}

We performed OS detection following the hybrid approach of Appendix~\ref{app:hybrid_approach} after executing Nmap and OpenVAS scans of each LAN.
Figure~\ref{fig:study_case_os_detection} shows the distribution of the different OS Families for each LAN, fot a total of 215 hosts.
Since we are interested in mapping each host to a VM Template, we need accuracy at the OS Generation level.
Excluding all entries belonging to the \emph{Other} Family (mostly network devices, firewalls, access points, etc.), we could unambiguously determine the OS Generation for about 60\% of the host virtual machines, while relying solely on Nmap we could map only about 25\% (mostly corresponding to Windows machines, for the reasons expressed in Section~\ref{sec:meth_ref_osdetection}, while with OpenVAS alone we reached 45\% overall. Unfortunately, due to problems with the KVM Hypervisor, we couldn't manage to emulate correctly the OSX Machines, and resorted to only keeping Windows and Linux VMs (the specific OS Generations will be detailed in the following).


  
\begin{figure*}
	\centering 
	\includegraphics[width=1.15\linewidth,trim={0.1cm 16.cm 0cm 0cm},clip]{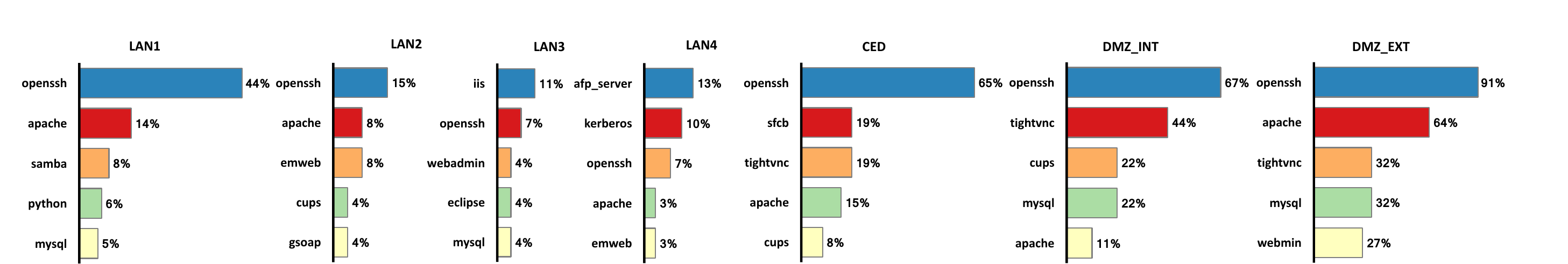}
	\caption{Top 5 installed services by percentage of installations in the various LAN of the reference network.}
	\label{fig:study_case_top_services}
\end{figure*}

OpenVAS and Nmap were also implied in the discovery of running services as well the known vulnerabilities for each host. 
Figure~\ref{fig:study_case_top_services} reports the top occurring services found in the various LANs.

\subsection{Physical infrastructure}
Table~\ref{tab:study_case_testbed_conf} reports hardware configuration of our Virtual Environment infrastructure.
\begin{table}[]
\centering
	\caption {Hardware Specifications}
	\begin{tabular}{|c|c|l|}
		\hline
		\multicolumn{1}{|c|}{\textbf{Type}} & \multicolumn{1}{c|}{\textbf{Name}}                                            & \textbf{Specifications}                                                                                                                                                                 \\ \hline
		Server                     & \begin{tabular}[c]{@{}c@{}}$2 \times$ Cisco UCS C240\\ M4S\end{tabular} & \begin{tabular}[c]{@{}l@{}}- $2 \times$ Intel Xeon E5-2620v3\\ (24 total threads)\\- 64 GB RAM\\ 
		- 8 x HDD 1.2 TB\end{tabular} \\ \hline
		\multirow{2}{*}{Switch}    & Cisco Catalyst 2960XR                                                & \begin{tabular}[c]{@{}l@{}}
		 - APM86392/512 MB RAM\\ - 48 x 1GB Ports \\\end{tabular}                                                                \\ \cline{2-3} 
		& Cisco Catalyst  WS-C3850                                             & \begin{tabular}[c]{@{}l@{}}
		 - MIPS/4 GB RAM\\- 24 x 10GB Ports\\\end{tabular}                                                                       \\ \hline
	\end{tabular}
	\label{tab:study_case_testbed_conf}
\end{table}
The two servers, connected via the L2 switch Cisco Catalyst 2960 form the backbone network for GlusterFS traffic, while the second L3 Cisco Catalyst WS-C3850 switch is used for the emulation environment (OVS bridges)
As detailed in the following, we will make use of 3 OVS bridges for networking.
Each bridge uses a physical network interface on each server. 
The basic templates for the VMs and the ISO Images for the various operating systems are saved in the Image Datastore. 
Since we couldn't count on high performance Ethernet interfaces on the nodes, we adopted the Linux bonding strategy to maximize the bandwidth on the backbone network. In particular, we used Linux kernel bonds with bond-mode 0, and 6 1GBps interfaces per node, achieving in this way a maximum bandwidth equal to 6 Gbps.

 
\subsection{Testbed Architecture}
\label{sec:studycase_testbed}

Due to lack of resources, and also considering other VMs necessary for networking, firewalling and generation of benign and malicious background traffic by external hosts, we could only afford to instantiate about half of the identified host machines.
We deployed all the VLANs necessary for replicating the reference network, using three OVS virtual bridges, namely \texttt{br1, br2 and br3}. The first bridge will be devoted to internal, business LANs, the second to service-related VLANs, and network devices (routers and a firewall), while the third bridge provides Internet connection to the whole testbed hosting the interfaces of all such machines that need direct internet connection.

For VLAN Routing at Layer3, we made use of three Virtual Routers, namely VR0, VR1 and VR2. Furthermore, as per the reference network, we use a Virtual Firewall, \emph{FW}. Virtual Routers run ZeroShell v3.8.2, a lightweight Linux Distribution suited 
for router emulation. For the firewall, we used PfSense 2.4.3.\\

\noindent More specifically, the architecture is divided in the following parts:\\
\textbf{Internal Area:} This area is composed by LAN1, LAN2, LAN3 and LAN4. The virtual networks are available to the virtual machines through the corresponding bridge \texttt{br1}. \\
\noindent	\textbf{DMZ Area:} CED and DMZ\_INT provide internal services to authorized hosts in the Internal Area through VR1. DMZ\_EXT contains servers exposing services to the Internet. Note that access to services in the DMZ area is allowed by well defined firewall rules specified on FW.\\
\noindent	\textbf{External Hosts:} This VLAN, as well as Attackers, is used to simulate a portion of the Internet, made by hosts that will connect to services from the DMZ\_EXT VLAN.\\
\noindent	\textbf{Attackers:} The Attackers network, hosts  malicious external attackers that want to exploit or attack hosted services (Section~\ref{sec:attack_scenarios}). 

DMZ Area, Attackers and External Hosts VLANs are deployed on bridge br2.
Table~\ref{tab:study_case_testbed_network} summarizes the list of VLANs, networks addresses and number of hosts, with installed operating systems and related public and private IPs and services installed in the service LANs. 
Note that both External Hosts, Attackers and servers in the DMZ\_EXT have ranges in the public IPv4 space: indeed we replicated a small portion of the internet inside our testbed. 

In order for the communication from such networks to correctly reach the DMZ\_EXT (instead of flowing towards the Externet VLAN), a static route was configured on VR2 to direct traffic with destination inside DMZ\_EXT through VR2's interface on FWVR2. 

Thus, traffic from such external hosts directed to DMZ\_EXT
will be routed by VR2, using the static route, and all the corresponding responses will return back to the originating hosts, since VR2 knows both External Hosts and Attackers through direct interfaces.\\ 

\begin{figure*}
	\centering 
	\includegraphics[width=1\linewidth]{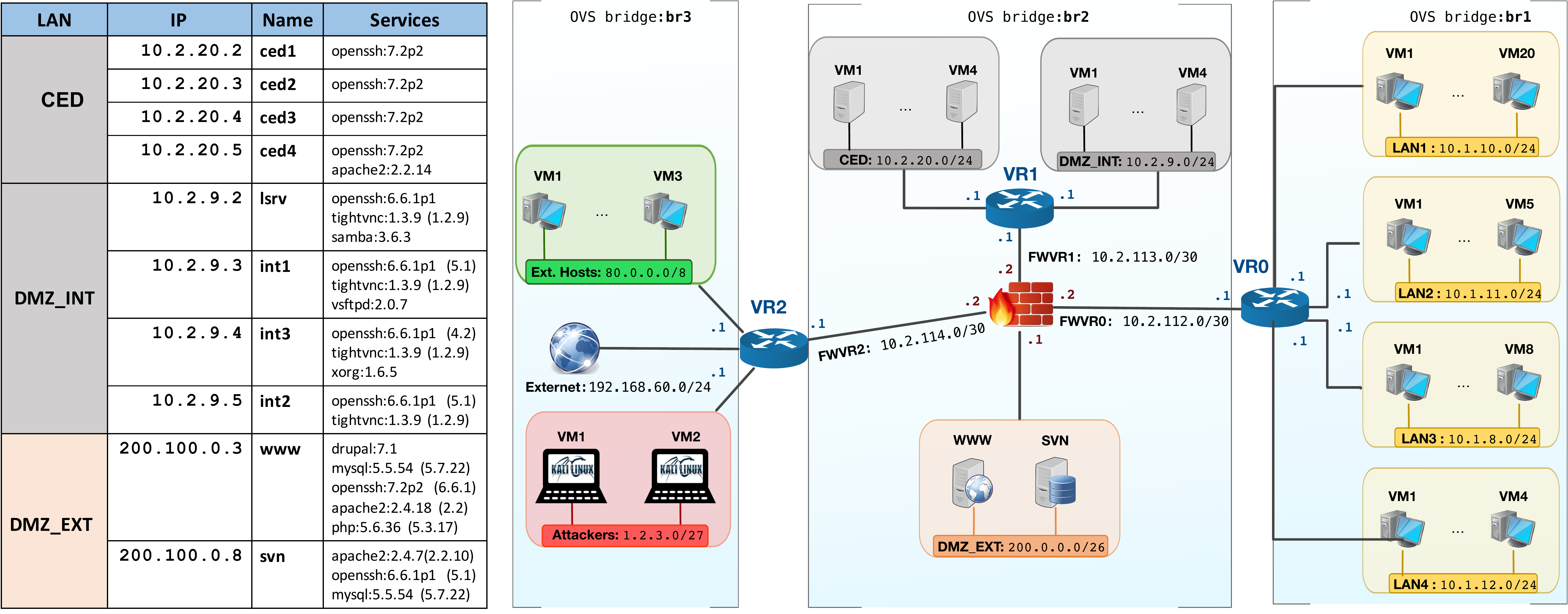}
	\caption{Testbed network architecture: The VLANs arrangement in L2 OVS bridges is highlighted using blue boxes. Installed services on service LANs are also shown. Reported package versions were automatically installed in the testbed, original ones from the reference network are in parentheses.}
	\label{fig:study_case_testbed_architecture}
\end{figure*}

\begin{table}
\centering
	\caption {LANs Operating Systems and IPs}
	\begin{tabular}{|c|c|l|l|}
		\hline
		\textbf{LAN}              & \textbf{Network}                                                                  & \textbf{OS}                                             & \textbf{IPs}                                                               \\ \hline
		\multirow{8}{*}{LAN1}     & \multirow{8}{*}{\begin{tabular}[c]{@{}c@{}}\textbf{10.1.10.}0/24\\ 20 hosts\end{tabular}}  & Ubuntu 12.04                                            & 16                                                                        \\ \cline{3-4} 
		&                                                                                   & Ubuntu 14.04                                            & 4, 22                                                                    \\ \cline{3-4} 
		&                                                                                   & Ubuntu 16.04                                            & \begin{tabular}[c]{@{}l@{}}6-10, 13\\15, 17, 19\end{tabular}      \\ \cline{3-4} 
		&                                                                                   & Ubuntu 16.10                                            & 18                                                                        \\ \cline{3-4} 
		&                                                                                   & Debian 7                                                & 12                                                                        \\ \cline{3-4} 
		&                                                                                   & Windows 10 (1803)                                   & 11, 14                                                                   \\ \cline{3-4} 
		&                                                                                   & Win. Server 2008 SP1                                 & 21                                                                        \\ \cline{3-4} 
		&                                                                                   & Win. Server 2012 R2                                  & 20                                                                        \\ \hline
		\multirow{4}{*}{LAN2}     & \multirow{4}{*}{\begin{tabular}[c]{@{}c@{}}\textbf{10.1.11.}0/24\\ 5 hosts\end{tabular}}   & Ubuntu 16.04                                            & .3                                                                         \\ \cline{3-4} 
		&                                                                                   & Windows 7                                               & 6                                                                         \\ \cline{3-4} 
		&                                                                                   & Windows 10 (1709)                                   & 5                                                                         \\ \cline{3-4} 
		&                                                                                   & Windows 10 (1803)                                   & 2, 4                                                                     \\ \hline
		\multirow{5}{*}{LAN3}     & \multirow{5}{*}{\begin{tabular}[c]{@{}c@{}}\textbf{10.1.8.}0/24\\ 8 hosts\end{tabular}}    & Ubuntu 14.04                                            & .6                                                                         \\ \cline{3-4} 
		&                                                                                   & Windows 7 SP1                                           & 2, 4                                                                     \\ \cline{3-4} 
		&                                                                                   & Windows 10 (1709)                                    & 5                                                                         \\ \cline{3-4} 
		&                                                                                   & Windows 10 (1803)                                   & \begin{tabular}[c]{@{}l@{}}3, 7, 9\end{tabular}                       \\ \cline{3-4} 
		&                                                                                   & Win. Server 2008 SP1                                 & 8                                                                         \\ \hline
		\multirow{4}{*}{LAN4}     & \multirow{4}{*}{\begin{tabular}[c]{@{}c@{}}\textbf{10.1.12.}0/24\\ 4 hosts\end{tabular}}   & Windows 7                                               & 2                                                                         \\ \cline{3-4} 
		&                                                                                   & \begin{tabular}[c]{@{}l@{}}Windows 7 SP1\end{tabular} & 3                                                                         \\ \cline{3-4} 
		&                                                                                   & Windows 10 (1709)                                    & 5                                                                         \\ \cline{3-4} 
		&                                                                                   & Windows 10 (1803)                                   & 4                                                                         \\ \hline
		\multirow{2}{*}{CED}      & \multirow{2}{*}{\begin{tabular}[c]{@{}c@{}}\textbf{10.2.20.}0/24\\ 4 hosts\end{tabular}}   & Ubuntu 10.04                                            & 5                                                                         \\ \cline{3-4} 
		&                                                                                   & Ubuntu 16.04                                            & 2-4                                                                       \\ \hline
		DMZ\_INT                  & \begin{tabular}[c]{@{}c@{}}\textbf{10.2.9.}0/24\\ 4 hosts\end{tabular}                     & OpenSUSE 11.4                                           & 2-5                                                                       \\ \hline
		\multirow{2}{*}{DMZ\_EXT} & \multirow{2}{*}{\begin{tabular}[c]{@{}c@{}}\textbf{200.100.0.}0/26\\ 2 hosts\end{tabular}} & Ubuntu 14.04                                            & 8                                                                         \\ \cline{3-4} 
		&                                                                                   & Ubuntu 16.04                                            & 3                                                                         \\ \hline
		Ext. Hosts                & \begin{tabular}[c]{@{}c@{}}\textbf{80.}0.0.0/8\\ 3 hosts\end{tabular}                      & Ubuntu 16.10                                            & \begin{tabular}[c]{@{}l@{}}20.40.2,\\ 100.3.50,\\ 20.55.21\end{tabular} \\ \hline
		Attackers                 & \begin{tabular}[c]{@{}c@{}}\textbf{1.2.3.}0/27\\ 2 hosts\end{tabular}                      & Kali Linux 4.15.0                                              & 2, 16                                                                    \\ \hline
	
	\end{tabular}
	\label{tab:study_case_testbed_network}
	\end{table}

\noindent \emph{Testbed deployment} The deployment of all VLANS, VMs, Virtual Routers and Firewall components was completely automated using a Python based module via the Python bindings of the XML-RPC APIs exposed by OpenNebula. 

In order to provide each machine with the appropriate services (with version) determined in the vulnerability surface identification phase, we created startup scripts according to each operating system family, that we feed with the services to be installed, so that they are fetched and installed automatically at first machine boot.

A particular effort must be devoted for the installation of the services, starting from the CPEs. Currently, there is no available data source that allows to match application CPEs to actual packages. 
For this reason the mapping must resort to either a manual association, or a tentative automatic association based on similarity. 

In particular, we resorted to manually mapping application CPE to software packages, that then were installed automatically using the closest version available in the repositories.

\section{Dataset}
As part of our work, we gathered a realistic dataset, comprising of benign and malicious traffic, that has been processed to obtain feature-rich labeled attack flows, useful for cyber security analyses such as training IDS and IPS classifiers and other machine-learning tasks, as well as for deep packet inspection investigation and related activities.
In particular, the following sections elaborate on our benign profile agents and the a number attack scenarios that we considered.

\subsection{Benign Traffic Agents}
\label{sec:dataset_generators}
To form the main part of our dataset, out aim was having realistic background traffic that comprises a number of widely used protocols, namely HTTP, HTTPS, SSH, SMB and SFTP, to act as background noise with respect to a number of traffic flows corresponding to specific instantiated attack patterns. To this end, we used a number of software agents written in Python to provide close-to-realistic user behavior for the VLANS from LAN1, through LAN4.\\
The agents' software consists in a number of traffic simulation jobs executed by worker threads managed by a scheduler, which feeds the threads following a specific \emph{traffic profile} shown in Table~\ref{tab:dataset_traffic_profile}), and assigned to each machine. In particular, on average, each job will be executed during the interval (Start, End) defined in the profile, a number of times depending on the Frequency profile.
The scheduler will randomize the time between the execution of each job to 
prevent the traffic from looking too synthetic.  
Furthermore, all jobs have an initial setup phase that always entails a degree of randomness in all configurable parameters, as well as multipliers for randomly delaying each request.\\

In the following, we summarize the behavior of the simulation jobs for the different protocols supported by our agents.\\\\
\textbf{HTTP/HTTPS generator:}  
An HTTP job, ``browses'' the internet starting at pre-defined root urls and simulates clicks by randomly following links on pages until a pre-defined click depth of 7 is met. Wait time between HTTP requests is chosen at random uniformly between 5 and 10 seconds.
    The number of total http requests is given by a random integer generated uniformly between 1 and of 20. The value of the parameters has been chosen following previous research~\cite{wang2004characterizing,white2007investigating}, that found that only about 5\% of sessions are longer than 20 requests and the average session length is about 7. Note that to generate request that are compatible with the associated OS, each agent will perform requests with a specific \texttt{User-Agent} that is automatically identified by the software based on the machine OS.\\\\
\textbf{SSH generator:} 
An SSH job connects to a host with the provided credentials using SSH and start sending commands selected at random from a list of supplied commands, keeping the connection open until a given period of time generated at random. 
The commands list is generated at startup time from a list of common commands such as \texttt{`ls', `cd', `cat /var/log/messages'}, etc.\\\\
\textbf{SMB generator:}
A SMB job connects to a SMB share with the provided credentials and uploads and downloads a number of files or directories. The files are randomly generated and their size are controlled by an optional size multilier of the default size of 8192 bytes.\\\\
\textbf{SFTP generator:} An SFTP generator connects to a host with the provided credentials using the SSH subsystem SFTP. It then starts uploading and downloading files provided with the same parameters of the SMB generator.

\begin{table}[]
\centering
\begin{tabular}{|c|r|r|r|c|c|llll}
\cline{1-6}
\textbf{Type} & \multicolumn{1}{c|}{\textbf{Start}} & \multicolumn{1}{c|}{\textbf{End}} & \multicolumn{1}{c|}{\textbf{Freq}} & \textbf{Hosts} & \textbf{Servers}       &  &  &  &  \\ \cline{1-6}
WEB           & 9:00                                & 18:30                             & 15                                 & All            & *                      &  &  &  &  \\ \cline{1-6}
SMB1          & 9:00                                & 18:00                             & 60                                 & All but h2     & lsrv                 &  &  &  &  \\ \cline{1-6}
SMB2          & 14:00                               & 18:00                             & 180                                & All but h2     & lsrv                 &  &  &  &  \\ \cline{1-6}
SSH3          & 9:00                                & 18:00                             & 60                                 & h1             & ced2                   &  &  &  &  \\ \cline{1-6}
SSH4          & 9:00                                & 18:00                             & 240                                & h2             & lsrv, ced1-3, int1-3 &  &  &  &  \\ \cline{1-6}
SSH5          & 9:00                                & 18:00                             & 60                                 & h3             & ced3                   &  &  &  &  \\ \cline{1-6}
SSH6          & 9:00                                & 18:00                             & 60                                 & h4             & int1                   &  &  &  &  \\ \cline{1-6}
SSH7          & 9:00                                & 18:00                             & 60                                 & h5             & int2                   &  &  &  &  \\ \cline{1-6}
SFTP          & 9:00                                & 18:00                             & 240                                & All but h2     & int3                   &  &  &  &  \\ \cline{1-6}
\end{tabular}
\caption{Traffic generator profiles for the various supported protocols. All, refers generically to all hosts in the four business VLAN1-4. The reported custom hosts correspond to: h1 (10.1.10.9), h2 (10.1.10.8), h3 (10.1.10.17), h4 (10.1.10.13), h5 (10.1.11.3).}
\label{tab:dataset_traffic_profile}
\end{table}


The agents' code has been deployed at machine instantiation time using init scripts, and the resulting processes are configured as Systemd (Ubuntu 14.04+, Debian) or Upstart Jobs (Ubuntu until 14.04), and using Task Scheduler on all Windows OS, configured to run as automatic startup jobs and to be automatically restarted in case of failures.

\subsection{Attack profiles and scenarios}
\label{sec:attack_scenarios}
To give a practical demonstration of the power and flexibility of our emulated environment, we performed various cyber attacks in the testbed, covering a diverse set of attack scenarios. In particular, starting from common attack taxonomies, 4 attack profiles were selected, and executed by using related tools and codes.\\\\
\textbf{Brute Force:}
A brute force attack can target different types of services. The main goal of this attack is obtaining private information, \ie a password, or private key, by testing many input combinations through a trial and error process, until the private information is finally gathered. This attack is one of the easiest to perform and detect, given the huge amount of traffic generated by the probing process. 
In particular, we performed a Dictionary-based brute-force attack to an OpenSSH daemon running on a Linux Machine.\\\\\
\textbf{Heartbleed:}
Heartbleed is a vulnerability of the OpenSSL implementation of the SSL protocol. The name comes from a specific function introduced in version 1.0.1, called \emph{HeartBeat}: this extension allows the clients to test and keep a connection alive without the need of re-negotiating it from scratch. Thus, suffering from a buffer-over-read vulnerability, it allows the clients to read more data than expected, possibly leaking sensitive information. To perform this attack, an SSL connection must be instantiated with a vulnerable server, sending a malformed Heartbeat request.\\\\
\textbf{Web Application Attack:}
Attacks on web applications may target different services. The OWASP Top 10~\footnote{\url{https://www.owasp.org/index.php/Top_10-2017_Top_10}} shows the most common pattern for this specific attack surface, where the targets span from simple websites to complex CMS. As a practical case, we targeted the Drupal CMS version 7.31, vulnerable to a Remote Command Execution (RCE) exploit called drupalgeddon2 (CVE 2018-7600). With this attack, it is possible to obtain control of the attacked server by exploiting a sanitization vulnerability in the AJAX requests 
allowing the attacker to force the vulnerable server to execute arbitrary malicious code.\\\\
\textbf{LAN Attack (+ Ransomware Deployment):}
This attack pattern was introduced to represent an insider threat, \ie a malicious user who has access to the organization's internal network. We conducted a lateral spreading attack by exploiting an SMB vulnerability (CVE-2017-0144) affecting various Microsoft Windows versions, as made by the Eternalblue exploit used in 2017 for spreading the DoublePulsar backdoor. In this case, the attack started from an intruder Kali Linux, connected to the LAN1. The attack begins with a series of Nmap scans (alive, services) towards LAN4, where a victim is found having a vulnerable Windows 7 system. The attacker installs a reverse TCP shell via Eternalblue and using a local HTTP server installed on the attacker machine downloads and runs the famous WannaCry software on the victim machine.\\

All the specific attack steps as well as the corresponding timeline will be publicly released along with the full capture dataset. 

\subsection{Data Collection}
Our data collection objective was gathering full traffic capture for all VLANs pertaining to the department network infrastructure, namely LAN1 through LAN4, FWVR0, FWVR1, as well as the service LANs CED, DMZ\_INT and DMZ\_EXT. 
\begin{table*}[]
\scriptsize
\centering
\begin{tabular}{|l|r|r|c|r|r|r|r|r|r|}
\hline
\multicolumn{1}{|c|}{\textbf{Name}} & \multicolumn{1}{c|}{\textbf{Attacker}} & \multicolumn{1}{c|}{\textbf{Victim}} & \textbf{Day} & \multicolumn{1}{c|}{\textbf{Start}} & \multicolumn{1}{c|}{\textbf{End}} & \multicolumn{1}{c|}{\textbf{Time (s)}} & \multicolumn{1}{c|}{\textbf{\# pkts}} & \multicolumn{1}{c|}{\textbf{Avg. pps}} & \multicolumn{1}{c|}{\textbf{Avg. size (B)}} \\ \hline
\textbf{Heartbleed}                 & 1.2.3.3                                & 200.100.0.8                          & \textbf{25}  & 15:01:51                            & 15:06:08                          & 256.3                                  & 144                                   & 0.6                                    & 720                                         \\ \hline
\textbf{Bruteforce}                 & 1.2.3.3                                & 200.100.0.3                          & \textbf{25}  & 16:03:37                            & 17:59:13                          & 6935.5                                 & 504641                                & 72.76                                  & 137                                         \\ \hline
\textbf{Web}                        & 1.2.3.16                               & 200.100.0.3                          & \textbf{27}  & 10:39:05                            & 11:06:50                          & 1665.2                                 & 554                                   & 0.33                                   & 581                                         \\ \hline
\textbf{LAN}                        & 10.1.10.2                              & 10.1.12.2                            & \textbf{27}  & 12:21:30                            & 12:40:57                          & 1167.5                                 & 2816                                  & 2.41                                   & 1464                                        \\ \hline
\textbf{PortScan}                   & 10.1.10.2                              & LAN4                                 & \textbf{27}  & 12:07:32                            & 12:21:30                          & 838                                    & 24388                                 & 29.10                                  & 61                                          \\ \hline
\end{tabular}

\caption{Time and statistics for the attack instances deployed in the testbed. Times are in  CEST timezone}
\label{tab:attacks_stats}
\end{table*}

Following the directions discussed in Section~\ref{sec:meth_data_coll}, to perform the full traffic capture, we deployed a sniffer VM for each pair $(n, b)$ (where $n$ is a node and $b$ is an OVS bridge) having one network interface on each VLAN $v$ belonging to bridge $b$. Each of such port has been configured to be the output port of a SPAN mirror, that receives traffic from all interfaces of VLAN $v$, belonging to machines deployed on node $n$. We thus captured traffic from all sniffers network interfaces using the Linux network traffic capturing tool \texttt{tcpdump}.


The data capturing period started at 14:30 on 25 July 2018 and continuously ran for an exact duration of 4 days, ending at 14:30 on 29 July 2018. 

\noindent As described in the previous section, it contains both benign traffic generated with our agents as well as malicious traffic flows. Concerning the latter, we deployed some instances of the attack profiles reported in Section~\ref{sec:attack_scenarios} and extracted the corresponding network traffic from the captures to obtain a list of attack-traffic subcaptures. Statistics for each attack are shown in Table~\ref{tab:attacks_stats}, \ie involved hosts, attack timespans (start, end and duration), total number of packets, average packets per second and average packet size in bytes.
The PortScan row refers to the Nmap scanning phase of the LAN Attack scenario.
Furthermore, we processed all attack-traffic captures with a flow-based feature extractor~\cite{lashkari2017characterization,sharafaldin2018toward}, in a way to extract 80 features for all included flows, where each flow corresponds to a \texttt{(SrcIP, SrcPort, DstIP, DstPort,Protocol)} tuple.
The extracted features are useful for training Intrusion Detection and Prevention Systems (IDS, IPS) classifiers and other machine-learning tasks for cyber security analyses~\cite{sharafaldin2018toward}.\\
\noindent To summarize, the dataset~\footnote{The full dataset is available at \url{https://www.cis.uniroma1.it/dataset_ICDCN2019}, released under the CC-BY license.} contains the following information:\\
\begin{itemize}
\item \textbf{Full daily captures:} One PCAP file per day with the full capture for each VLAN.
\item \textbf{Attacks:} A separated PCAP files for each attack, as per Table~\ref{tab:attacks_stats}, a well as a description of the necessary steps to replicate it.
\item \textbf{Benign:} A separate PCAP file with all traffic from LAN1-LAN2 for 26/07, containing only benign (agent-related) flows, as per Table~\ref{tab:dataset_traffic_profile}).
\item \textbf{Features:} for each PCAP file, a CSV files containing, for each extracted flow, the values for 80 traffic related features.
\end{itemize}

\section{Conclusions and Future Work}
This paper proposes a methodology for the design, development and deployment of cloud-based emulation environments.
Based on our practical experience, we provided some recommendation about technological choices to support the creation of virtual environments.
We also show the effectiveness of our approach by applying our methodology to build an emulation environment for the study case of our Department network.
Finally, we show the flexibility of the deployed environment to simulate actual cyber attacks and benign traffic, that we collected in a publicly released dataset containing complete network traces, enriched with labeled features.
A limitation of our work is the lack of an evaluation of the accuracy of the information acquired from the reference network mapping phase, due to lack of precise ground truth, which would require an intrusive approach. 
As a future step in this direction, we plan to work on the definition of a strategy to measure the affinity of the final emulated environment and the original reference network. 
The next step in our investigation will be toward the definition of novel approaches to develop user behavior profilers to improve the quality of benign data generators in terms of closeness to real human behavior.
Additionally, we plan to investigate approaches to fully automate the installation of services, which is currently an open issue.


\section*{Acknowledgements}
This work has been partially supported by CINI Cybersecurity National Laboratory within the project FilieraSicura: Securing the Supply Chain of Domestic Critical Infrastructures from Cyber Attacks (\url{www.filierasicura.it}) funded by CISCO Systems Inc., Leonardo SpA and by the INOCS Sapienza Ateneo 2017 Project (protocol number RM11715C816CE4CB). The authors are deeply grateful to Tiziana Toni for her precious support and guidance, and Prof. Antonio Cianfrani for useful discussions.

\bibliographystyle{abbrv}
\bibliography{bibliography}
\newpage
\appendix
\section{IaaS Comparison: OpenStack vs OpenNebula}
\label{app:iaas_comp}
Started as a research project back in in 2005, OpenNebula~\cite{milojivcic2011opennebula} provides a simple-to use but feature-rich and flexible solution for the comprehensive management of virtualized data centers to enable private, public and hybrid IaaS clouds~\cite{von2012comparison}. Storage, virtual machine, data transfer, network management, and job scheduler are configured by OpenNebula services managementIt allows integration with different storage and network infrastructure configurations, and hypervisor technologies.

OpenStack~\cite{sefraoui2012openstack} first appeared in July 2010, and consists in a set of software tools for building and managing cloud computing platforms for public and private clouds. It builds on API stacks that communicate using a queuing service. It is composed of around forty individual components available. The storage may be deployed in different ways and stored locally or distributed. 

Next, we  provide a short discussion about the major differences between them.
\paragraph{Internal Organization.} While OpenStack comprises many different subprojects, among which: Orchestrator, Ceilometer (metering), swift (object storage support), Neutron (networking), Keystone and hybrid support aimed at building the different subsystems in a cloud infrastructure (each of them with its own API interface, and separate configuration effort), while OpenNebula offers a single integrated, comprehensive management platform for all cloud subsystems, with a single XML-RPC API interface..
\paragraph{Software Deployment.} An important feature is the easiness of installation. From our own experience OpenNebula was found easier to deploy than OpenStack because only a machine must be configured as front-end, which will host the cloud manage, and all the other machines will be slave nodes, thus only requiring the hypervisor daemon. OpenStack is more difficult because it requires the configuration of its multiple components according to a specific configuration~\cite{von2012comparison}.
\paragraph{Storage.} Storage is very important in cloud computing since all the images have to be managed and available for users anytime. OpenStack provides a sophisticated storage system called Swift and the images are transferred using ssh or http/s. OpenNebula does not provide a cloud storage product, and natively supportsfrom non-shared file systems with image transferring via SSH to shared file systems (NFS, GlusterFS, Lustre) or LVM with CoW (copy-on-write), and any storage server, from using commodity hardware to enterprise-grade solutions. By default it uses a shared file system, typically NFS or GlusterFS to store all files (see also Section~\ref{sec:meth_env_dfs}). 
\paragraph{Networking.} Both frameworks provide different options to manage the network. OpenStack manages the networks in two modes: flat networking and VLAN networking where the first one uses Ethernet bridging to connect multiple compute hosts together. In OpenNebula the networks can be defined to support VLAN tagging that requires support from the hardware switches and Open vSwitch to restrict network access with Open vSwitch.
\paragraph{Hypervisors.} Both of them support KVM, VMWare and Xen, which are the most popular open source hypervisors. Additionally, OpenStack also supports LXC, HyperV and Xen Server.\\
\paragraph{Open Source model and Governance} Both projects release code under the liberal Apache 2.0 license, follow a transparent development process with a public roadmap, and have the same license agreement for new contributions. While OpenStack is controlled by a Foundation~\footnote{\url{https://www.openstack.org/foundation/companies/}} driven by vendors that also sell their enterprise-grade proprietary implementations of its subcomponents, OpenNebula instead only release a single, free, community level dstribution, and is managed by a single organization backed by a community of developers.

We refer the interested reader to~\cite{vogel2016private} where the authors perform a deep comparison of the two projects, from the point of view of flexibility, resiliency and performance.
Overall, enterprise networks can benefit from OpenStack's proprietary implementations by HP, Red Hat and IBM that provide extended features, custom enhancements and integrations, as well as paid customer support. However, this can also translate in lock-in phenomena that erode compatibility and interoperability.
\section{OS Detection: a Hybrid Approach}
\label{app:hybrid_approach}
\begin{figure}[H]
	\centering 
	\includegraphics[height=2cm, width=8cm]{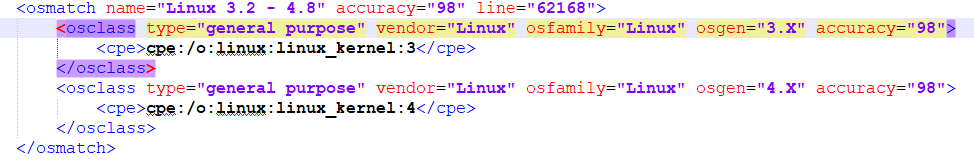}
	\caption{OS features}
	\label{fig:meth_ref_osmatch}
\end{figure}

Nmap's algorithm for detecting matches is relatively simple. It takes a subject fingerprint and tests it against every single reference fingerprint in \textit{Nmap-os-db}, and produces a list of \emph{OS matches} (see Figure~\ref{fig:meth_ref_osmatch}). When there are no perfect matches, Nmap adds an accuracy percentage.\\ 
The same OS release may fingerprint differently based on what network drivers are in use, user-configurable options, patch levels, processor architecture, amount of RAM available, firewall settings, and more. Sometimes the fingerprints differ for no discernible reason \cite{richardson2010limits}.
The probes and response matches are located in the \textit{Nmap-os-db} file. Nmap will attempt to identify the following parameters:
\begin{itemize}
	\item \textbf{Vendor Name} The vendor of the OS such as Microsoft or Sun.
	\item \textbf{Operating System} The underlying OS such as Windows, Mac OS X, Solaris.
	\item \textbf{OS Generation} The version of the OS such as Vista, XP, 2003, 10.5, or 10.
	\item \textbf{Device Type} The type of device such as general purpose, print server, media, router, WAP, or power device.
\end{itemize}
OS detection is enabled by providing the -O parameter. Additionally, one can increase the verbosity level with -v for even more OS-related details. Nmap includes several command-line options to configure the OS detection engine. If Nmap can't make a perfect match for an OS it will guess something that is close, but not exact. To make Nmap guess more aggressively, it can be used the \textit{--osscan-guess} command-line option.\\
The main problem with Nmap OS Detection concerns linux operating systems: given that the os fingerprinting technique mostly relies on network traffic features, the only observable differences are determined by different kernel levels, hence the impossibility of Nmap to discriminate between different Linux distributions.
For our mapping purposes (OS and services), we exploited a combination of Nmap and OpenVAS, which we describe next.

Most os fingerprints, also have a Common Platform Enumeration (CPE) representation, like \textit{cpe:/o:linux:linux\_kernel:2.6}. Common Platform Enumeration (CPE) is a structured naming scheme for information technology systems, software, and packages. Based upon the generic syntax for Uniform Resource Identifiers (URI), CPE includes a formal name format, a method for checking names against a system, and a description format for binding text and tests to a name.
Sometimes Nmap gives more than one CPE for a specific \textit{osmatch}, including the name, the type, the OS family and the accuracy. A good strategy is to take all those that have the highest accuracy for each IP. Thus, in some cases, we can have one or more CPE for an \textit{osmatch}.

However, for the emulation environment the results of Nmap are not sufficient, since specifically we need the correct OS generation for each host and not only the kernel linux version. OpenVAS tool uses different scanning tools and it provides among other functionalities, a good OS Detection. In particular, OS Detection Consolidation and Reporting consolidates the OS information detected by several NVTs and tries to find the best matching OS. 

For these reasons, for our OS detection task, we decided for a hybrid strategy that merges information from these two tools. Summarized in Figure~\ref{fig:meth_ref_osmatch}. 
\begin{figure}
	\centering 
	\includegraphics[width=1\linewidth]{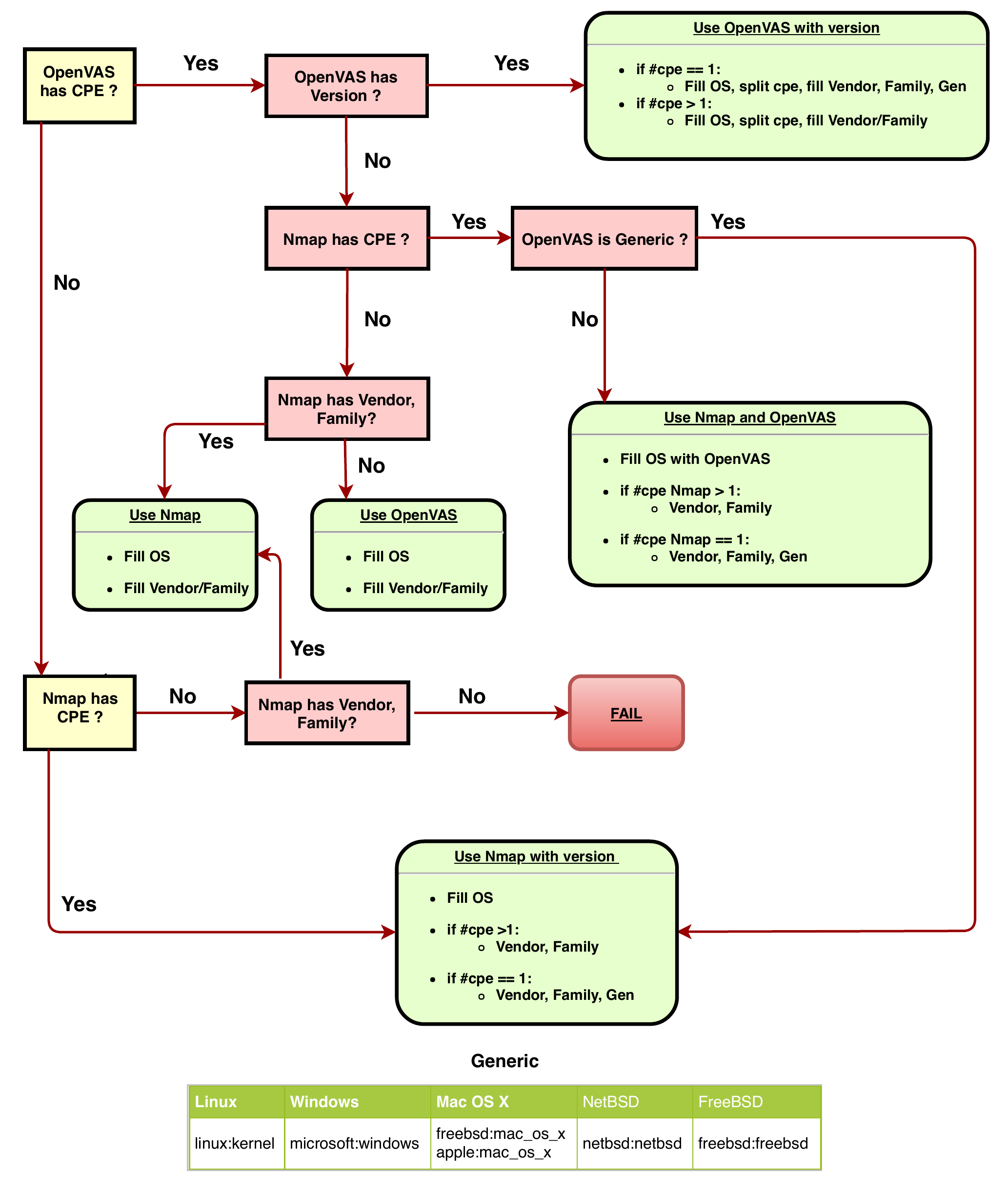}
	\caption{OS Matching Flow}
	\label{fig:matching_os}	
\end{figure}
The algorithm takes as input the output from Nmap and OpenVAS and for each entry, the network host, identifies the OS, Vendor, Family and Gen like \textit{Linux, Linux, Ubuntu, 16.04} for Linux hosts and \textit{Windows, Microsoft, 7, SP1} for Windows hosts. In this part was not considered the Microsoft Windows XP as operating system since it is a largely not used anymore and probably it can be a false positive.
For best results, the algorithm starts with considering the output from OpenVAS since it is more accurate with the Linux distributions. In particular, when it matches a Linux family it can provide also the name of the distribution and its version. In the case tools have more than one CPE for a specific host the algorithm will not provide the OS generation. The following are all the possible outputs:
\begin{itemize}
	\item \textbf{Use OpenVAS with version:} This output takes in consideration only the output from OpenVAS. In particular, for each entry it checks if the CPE is present. If yes, the OpenVAS output will be considered.
	\item \textbf{Use Nmap with version:} If an entry has not a CPE from OpenVAS the algorithm considers the Nmap output if the last one has a CPE and if the CPE of OpenVAS is generic, that is belongs from too general categories. 
	\item \textbf{Use OpenVAS and OpenVAS:} In this part the algorithm use the OpenVAS output for the OS, since it is not general, and Nmap output for the OS family and generation.
	\item \textbf{Use OpenVAS:} This kind of output and the last one are less accurate than previous ones since the algorithm reaches these states when neither Nmap nor OpenVAS return a valid CPE for an entry. The output brings only information about the OS, Family and Vendor but nothing about the generation. 
	\item \textbf{Use Nmap:} This part produces the same output as the precedent one using only the information from Nmap.
	\item \textbf{Fail:} Sometimes, for some entry like routers or firewall there are not CPE and not even information for the OS family and vendor. In these rare cases, the algorithm conclude with a fail status. To overcome this case, a manual checking could be necessary.
\end{itemize}

\end{document}